# Transformer-based prediction of two-dimensional material electronic properties under elastic strain engineering

Haoran Ma[1], Yuchen Zheng[1], Leining Zhang[2], Xiaofei Chen[1], Dan Wang[1*]

[1]School of Materials Science and Engineering, Beijing Institute of Technology, Beijing 100081, China

[2]School of Chemistry and Chemical Engineering, Beijing Institute of Technology, Beijing 100081, China

*Corresponding author. E-mail: wangdan7217@hotmail.com

**Abstract:**

Strain engineering provides a powerful route for tuning the electronic properties of two-dimensional (2D) materials, but exploring the full multidimensional strain space with density functional theory (DFT) is computationally prohibitive due to the nonlinear coupling between normal and shear components. In this work, we introduce a Transformer-based, multi-target surrogate model framework that achieves DFT-level bandgap prediction accuracy, reaching a mean absolute error of 0.0103 eV while retaining full interpretability through attention-weight analysis. The learned self-attention map consistently identifies shear strain as the interaction center that influences both bandgap and phonon stability, an insight not readily captured by classical feature-importance metrics. This work establishes attention-based architectures as physically interpretable surrogate models for multi-property prediction, offering a generalizable strategy for accelerating deep elastic strain engineering in materials informatics.

**Introduction:**

Two-dimensional (2D) materials have emerged as promising candidates for next-generation electronic and optoelectronic devices due to their unique electronic properties, mechanical flexibility, and quantum confinement effects[1,2]. Strain engineering provides a powerful route to band-structure control, because deliberate mechanical deformation adjusts bandgap, carrier mobility, and optical absorption while leaving the material chemistry unchanged[3–6]. For 2D materials, the strain space itself spans three in-plane normal ($\varepsilon_{xx}$, $\varepsilon_{yy}$) and shear ($\varepsilon_{xy}$) components, which can be varied independently, and it grows combinatorially with the number of sampled points[7]. Because the resulting search space expands exponentially, exhaustive experimental or first-principles exploration quickly becomes impractical and often leaves optimal strain-engineered configurations undiscovered[8].

While first principles density functional theory (DFT) offers accurate property predictions, its computational cost prevents systematic exploration of this vast strain space. Carrying out DFT for the thousands of strain combinations required in a multi-dimensional setting is computationally prohibitive[9]. This bottleneck has catalyzed the adoption of machine learning (ML) approaches, which offer orders-of-magnitude speedup while maintaining reasonable accuracy[10,11]. However, current state-of-the-art frameworks face specific limitations. For instance, modern strain-property models such as those by Shi et al[7] and Tsymbalov et al[12] often require volumetric band-structure representations (full 3D energy grids across the Brillouin zone) as inputs. This reliance necessitates extensive, costly first-principles calculations for data preparation, partially negating the efficiency gains of the ML surrogate.

Beyond computational cost, interpretability remains critical challenge. The approach of Tsymbalov et al[12] has demonstrated the power of convolutional neural networks (CNNs) for strain-engineering of semiconductor band structures, yet understanding how individual strain-tensor components contribute to property changes remains difficult due to the "black box" nature of deep CNNs[12]. Conversely, classical tree-based ensemble methods, such as random forests and gradient boosting, offer superior computational efficiency[13] but treat features largely independently[14]. This independence arises from their fundamental architecture[14]: decision trees split data based on single feature values at each node, processing features sequentially rather than modeling explicit interactions. While ensemble methods can indirectly capture some dependencies through hierarchical splits, these interactions remain implicit and local, making it difficult to quantify cooperative effects. Consequently, it is difficult for these methods to capture the nonlinear coupling between normal and shear strain components that is essential for accurate strain-property mapping[15]. Feedforward neural networks (FNNs) can model nonlinear relationships[16] but lack mechanisms to explicitly quantify feature interactions, leaving the learned representations opaque[17]. These limitations become particularly critical in strain engineering, where the cooperative effects of $\varepsilon_{xx}$, $\varepsilon_{yy}$, and $\varepsilon_{xy}$ determine material response.

To bridge the gap between predictive accuracy and physical interpretability, self-attention mechanisms, powered by Transformer architecture, offer a potential solution to address these limitations by capturing complex feature interactions in materials[18]. Unlike conventional architectures, Transformers learn feature relationships directly by weighting query–key similarities, simultaneously achieving high precision and tranparency[18,19]. Their multi-head attention aggregates information from several complementary perspectives, allowing the model to capture how strain components interact to shape material response. Because the attention weights themselves quantify each interaction, the framework provides enhanced interpretability, offering a direct window into the underlying physics consistent with recent successes in other high-dimensional scientific domains[20]. By operating directly on strain tensor components, an attention-based Transformer effectively bypasses the need for expensive volumetric band-structure data preparation.

In this work, we present a Transformer-based ML framework combining Latin Hypercube Sampling with self-attention mechanisms for high-throughput prediction of 2D material properties under strain engineering. We achieve DFT-level accuracy for bandgap prediction, yet retain interpretability through attention-weight analysis. The self-attention mechanism pinpoints shear strain $\varepsilon_{xy}$ as the interaction hub linking bandgap modulation and stability, complementing what classical feature-importance metrics reveal. This combination of computational efficiency, accuracy, and physical insight enables rapid screening of strain-engineered 2D materials.

# Results & Discussion:

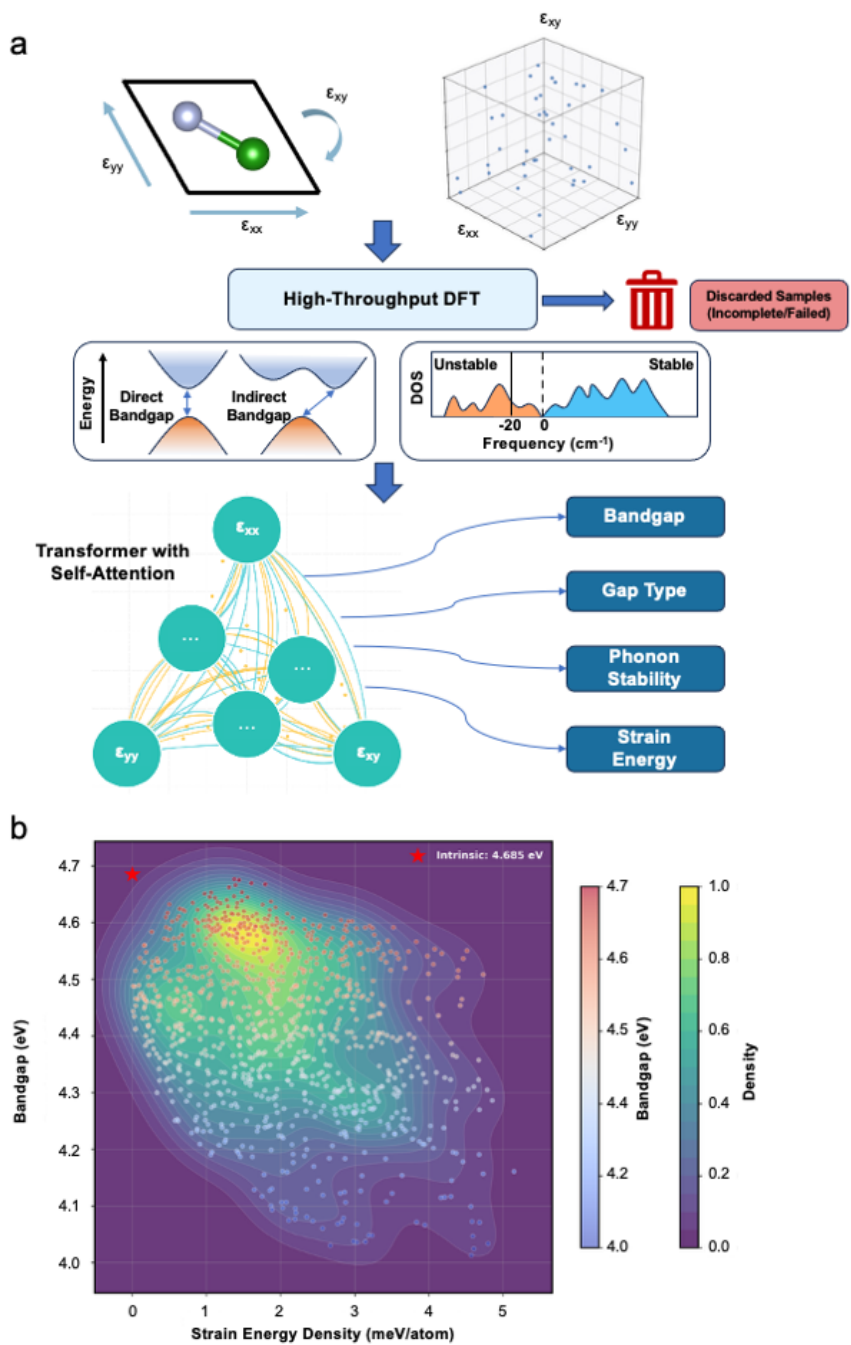

**Figure 1. High-throughput computational workflow and strain-property visualization**
a) Automated strain-to-property workflow combining Latin Hypercube Sampling, DFT, and phonon analysis. b) Density plot mapping strain energy density versus bandgap, with green-to-yellow contours indicating data density (equally spaced colorbar), highlighting regions of high data concentration

## Dataset Characterization and Computational Workflow

The bandgap is a fundamental electronic property that directly determines a material's optoelectronic behavior, including light absorption, carrier generation, and device efficiency. For 2D materials under strain engineering, precise bandgap control enables targeted design of photodetectors, light-emitting diodes, and solar cells with desired spectral responses. Moreover, the bandgap serves as a sensitive probe of electronic structure changes, making it an ideal property for validating strain-property relationships and understanding how mechanical deformation modulates quantum confinement effects in low-dimensional systems.

To systematically capture these dependencies, we developed an integrated computational and ML pipeline, as illustrated in Figure 1a. In this workflow, Latin Hypercube Sampling tiles the strain space, automated DFT and phonon calculations populate the database, and downstream ML models—including classical algorithms, feedforward neural networks, and Transformers—learn the strain-to-property relationships. Each case advances through structure relaxation, electronic-structure evaluation, phonon stability checks, and property extraction, yielding bandgap, stability, strain energy density, and direct-gap indicators without manual intervention. The fully automated loop transforms raw strain parameters into a multi-target dataset ready for high-throughput screening and model development.

Based on this comprehensive dataset, Figure 1b visualizes the quantitative bandgap tuning achieved via strain engineering. The relationship between strain energy density and bandgap modulation across the entire dataset is revealed in the density visualization shown in Figure 1b. Green-to-yellow density contours highlight regions of high data concentration, with a prominent peak centered around 4.5 eV bandgap and 1-2 meV/atom strain energy density, demonstrates the smooth and continuous nature of the strain-property correlation landscape. The density colorbar uses equally spaced intervals for uniform visualization of data distribution. Most stable configurations cluster in moderate strain and energy regions. The intrinsic 4.685 eV bandgap at zero strain (red star) serves as a reference for strain-induced shifts; roughly 85% of strained configurations fall below this value, with reductions between 0.01 and 0.67 eV. Kernel density estimation smooths discrete samples into continuous contours, making nonlinear trends immediately visible.

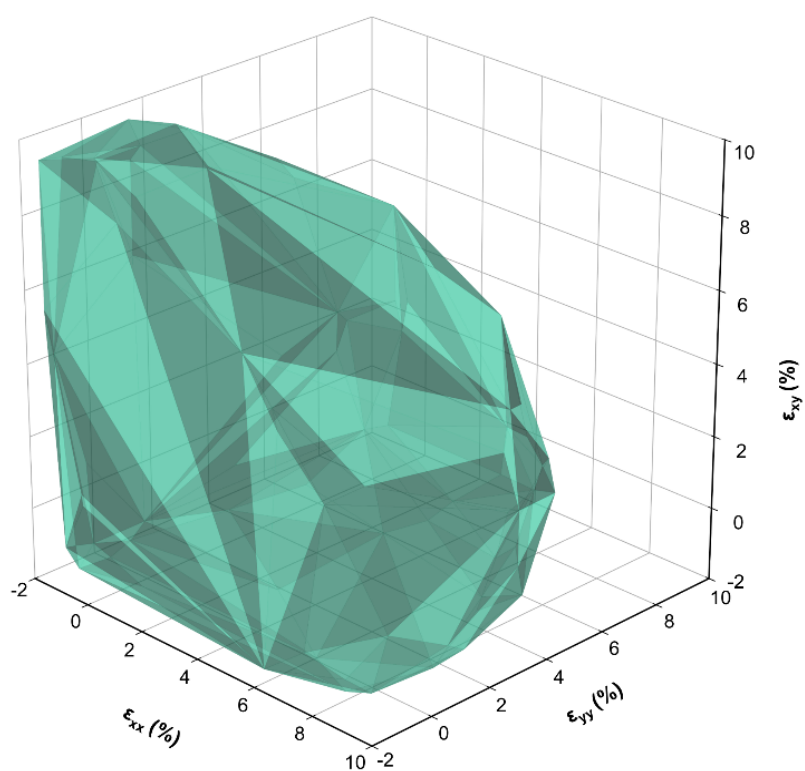

**Figure 2. Phonon stability landscape**

Three-dimensional convex hull representation showing 55.8% stable and 44.2% unstable regions in strain space, with the semi-transparent teal-colored polyhedral surface delineating the stable region boundary.

**Phonon Stability Landscape Analysis**

Regarding the statistical properties of the dataset, the bandgap distribution across the sampled strain configurations follows a well-behaved statistical pattern, with detailed statistical analysis provided in Figure S1 (Supplementary Information).

To determine the structural limits of the material, Figure 2 visualizes the phonon stability landscape across the strain space using a three-dimensional convex hull representation, shown from four complementary isometric perspectives (additional viewing angles are provided in Figure S2). The 992 successfully calculated configurations from 1000 sampled points are used to construct a convex polyhedron that encloses all stable configurations, with the semi-transparent teal-colored surface delineating the stable region boundary. The convex hull, computed from 554 stable points, forms a closed polyhedral surface composed of 162 triangular facets, effectively mapping the outermost boundary of the stable region in the three-dimensional strain space ($\varepsilon_{xx}$, $\varepsilon_{yy}$, $\varepsilon_{xy}$). Stable regions form continuous, interconnected volumes, while unstable regions appear as isolated pockets and extended peripheral domains outside the convex hull. Materials remain stable near the zero-strain origin, with stability probability decreasing as strain magnitudes increase. The visualization reveals anisotropic stability: stability boundaries exhibit complex geometric shapes, indicating that different strain component combinations produce different stability outcomes. Specifically, configurations dominated by normal strains ($\varepsilon_{xx}$, $\varepsilon_{yy}$) exhibit different stability patterns compared to those with significant shear components ($\varepsilon_{xy}$), identifying "safe operating regions" for strain engineering.

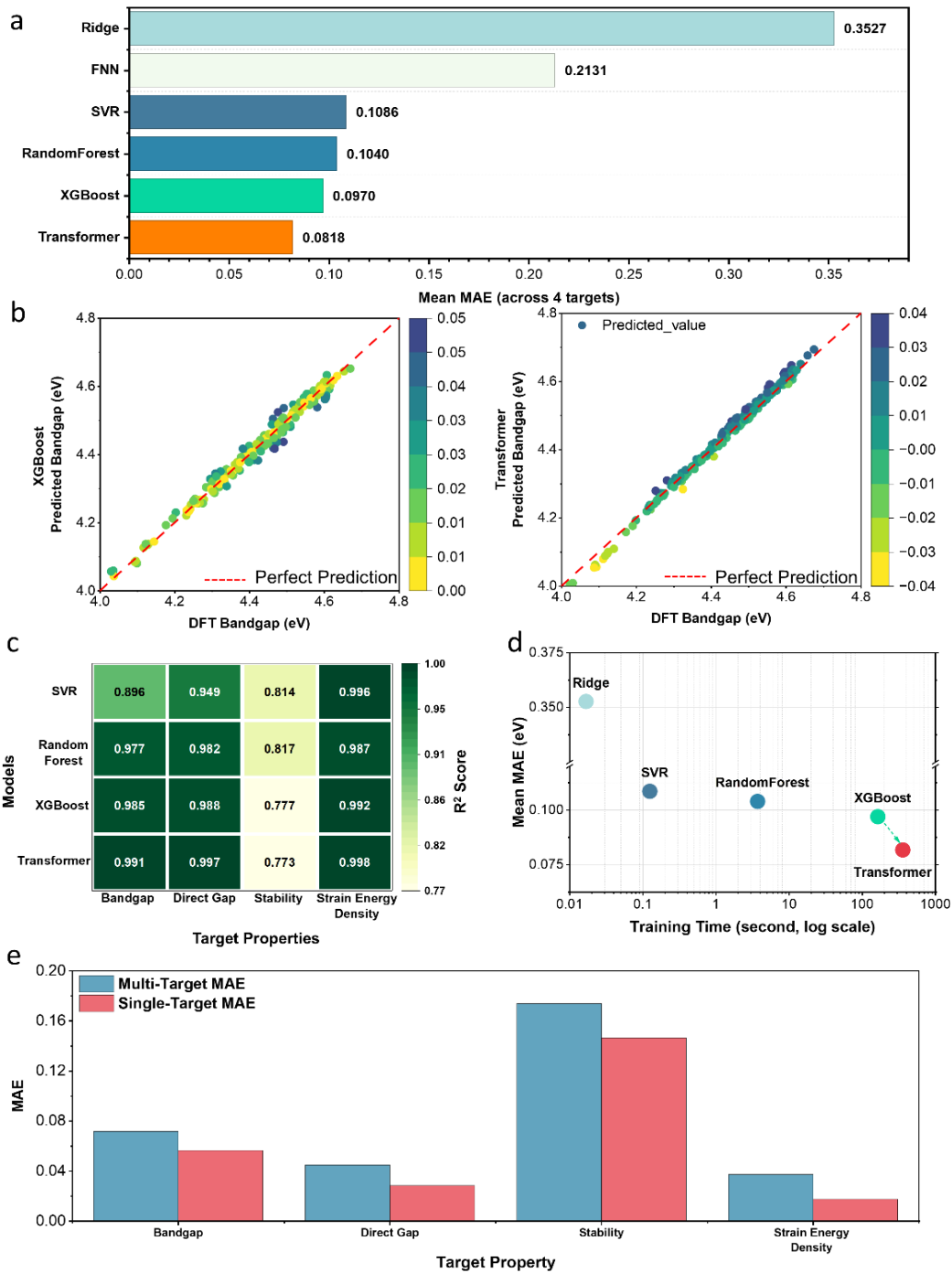

**Figure 3. ML model performance comparison and validation:** a) MAE of six models, highlighting the advantage of Transformer in accuracy and XGBoost in efficiency. b) Predicted versus DFT bandgap for XGBoost (left) and Transformer (right). c) Heatmap of R² across four targets for each model. d) Mean MAE versus training time; XGBoost lies on the Pareto frontier, while Transformer offers the highest accuracy. e) Performance comparison between multi-target and single-target models.

**Model Performance Comparison and Validation**

To benchmark the predictive capabilities of our framework, Figure 3a compares the performance across six distinct ML algorithms. Transformer achieved the best overall performance with MAE = 0.0818 eV (4-target) across all target properties, demonstrating superior capability in multi-target prediction tasks. XGBoost shows competitive performance with MAE = 0.0970 eV. We focus our detailed analysis on Transformer, XGBoost, Random Forest, and SVR in subsequent panels, as these models represent the best-performing architectures. FNNs show inferior performance (MAE = 0.2131 eV) and are therefore not discussed in detail.

Figure 3b provides a more granular view of prediction accuracy distribution, presenting scatter plots comparing predicted bandgaps against DFT-calculated values. Transformer demonstrated superior performance with MAE = 0.0103 eV, RMSE = 0.0136 eV, and $R^2$ = 0.991. XGBoost achieved DFT-level accuracy with MAE = 0.0135 eV and $R^2$ = 0.985. Both error distributions sit comfortably within the 10-20 meV energy uncertainty typical for plain DFT.

Expanding the evaluation to a multi-target performance analysis, the heatmap in Figure 3c compares $R^2$ values across four target properties. Transformer achieves superior performance for bandgap ($R^2 = 0.991$), direct gap ($R^2 = 0.997$), and strain energy density ($R^2 = 0.998$), while XGBoost performs best for stability prediction ($R^2 = 0.777$). To further quantify the prediction fidelity beyond the bandgap results, the Transformer model achieves a direct-gap indicator MAE of 0.0172 eV, a strain energy density MAE of 0.079 eV, and stability classification MAE of 0.186 (normalized scale). Overall, both models sustain $R^2$ values above 0.77, underscoring their ability to support simultaneous, multi-property screening.

Beyond pure accuracy, we also assessed training efficiency and computational cost. Figure 3d illustrates the trade-off between training time and prediction accuracy. The differences are substantial: Transformer requires roughly twice as long as XGBoost to reach the same error level. This reflects architectural differences: XGBoost's gradient boosting enables rapid convergence, while Transformer's multi-head attention layers incur higher computational costs to model high-order interactions. However, this cost is offset by the data-preparation advantage: unlike CNNs requiring volumetric inputs, our Transformer operates directly on strain tensors, avoiding extensive DFT band-structure calculations.

Finally, we conducted an ablation study to validate the architectural choice of multi-output prediction against single-target models. Figure 3e presents the ablation study comparing our multi-target Transformer with single-target models. Single-target models achieve better accuracy for individual properties (MAE improvements ranging from 18.6% to 113.3%). This is consistent with multi-task learning literature, where negative transfer can occur between competing tasks depending on task-relatedness and capacity[21,22]. For instance, bandgap and strain energy density exhibit a strong negative correlation ($r = -0.625$, see Figure S3), which may interfere with joint optimization. Nevertheless, the multi-target model offers a vital practical advantage: it requires training only one model instead of four, enabling rapid simultaneous screening while maintaining competitive accuracy ($R^2 > 0.99$ for most targets).

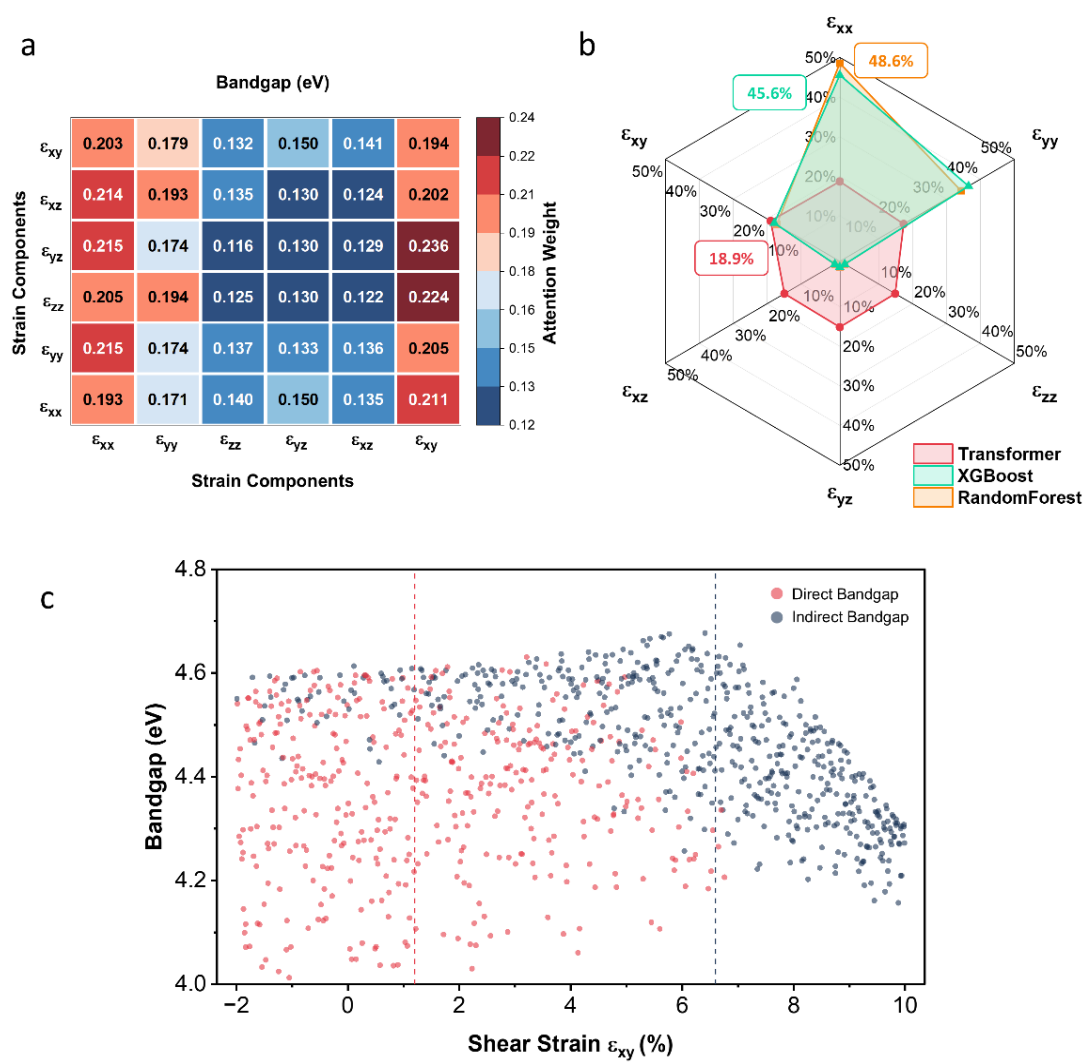

**Figure 4. Feature importance analysis and Transformer architecture**
a) Transformer attention heatmap revealing $\varepsilon_{xy}$ as the interaction center. b) Comparison of classical feature importance versus Transformer attention. c) Scatter plot of bandgap versus shear strain $\varepsilon_{xy}$, colored by bandgap type; vertical dashed lines indicate median $\varepsilon_{xy}$ values (direct: 1.2%, indirect: 6.6%).

## Feature Importance Analysis and Interpretability

To interpret the learned physics, the Transformer's self-attention mechanism offers fine-grained visibility into how strain components interact within the learned representation, as shown in Figure 4a. The 6×6 attention weight matrix averages eight heads across four encoder layers over 199 test samples, with brighter cells indicating stronger coupling between feature pairs. A consistent motif emerges: shear strain $\varepsilon_{xy}$ acts as the interaction hub of the strain–property network, receiving the largest inflows from $\varepsilon_{yz} \rightarrow \varepsilon_{xy}$ (23.6%) and $\varepsilon_{zz} \rightarrow \varepsilon_{xy}$ (22.4%), followed closely by $\varepsilon_{xx} \rightarrow \varepsilon_{xy}$ (21.1%). This attention pattern reflects the underlying physics of strain coupling: in 2D materials, shear deformation breaks rotational symmetry and couples normal strain components through geometric constraints, fundamentally altering the electronic structure in ways that linear elasticity models cannot capture. The eight-head architecture separates attention across complementary motifs: some heads emphasize interactions between two normal strains, others capture normal-shear couplings, and additional heads highlight higher-order effects. This layered perspective makes the attention view a natural complement to traditional feature-importance metrics and supplies interpretability that CNN-based strain models[12] cannot readily supply.

Complementing this attention-based perspective, the comparative analysis in Figure 4b highlights how classical ML and Transformer attention provide mutually reinforcing views of the strain–property landscape. Tree-based models treat features as largely independent and evaluate their marginal contributions through sequential splits; accordingly, they rank $\varepsilon_{xx}$ as the dominant driver (XGBoost: 45.6%, Random Forest: 48.6%) with $\varepsilon_{yy}$ following close behind. Transformer attention, by contrast, distributes weights more evenly across component pairs, assigning the highest interaction intensity to $\varepsilon_{xy}$ (18.9%) while keeping $\varepsilon_{xx}$ (18.7%) and $\varepsilon_{yy}$ (17.4%) in the same range. This divergence reflects the underlying modeling philosophy: classical methods quantify which single feature most reduces error, whereas self-attention captures how features interact within the learned embedding space, enabling the model to learn shared representations useful for predicting multiple properties simultaneously. Together, the viewpoints remain complementary: classical importance isolates dominant single-feature effects, whereas attention reveals features acting in concert, providing more diagnostic information than the aggregate maps from CNN baselines.

The physical consequence of this shear-strain dominance is visualized in Figure 4c, which shows how shear strain steers the direct/indirect bandgap transition. The scatter plot separates direct-gap (red) and indirect-gap (blue) configurations as a function of $\varepsilon_{xy}$. Direct-gap points cluster near the origin, whereas indirect-gap points extend toward larger positive shear. This distribution aligns with the interaction-hub role identified in Figure 4a: the wide dispersion of indirect-gap configurations at high shear strains empirically confirms that $\varepsilon_{xy}$ actively modulates band-edge ordering through symmetry breaking. This coupling effect is explicitly quantified by the high attention weights (calculated via Eq. 2), providing a direct mathematical link between the learned representations and the underlying lattice physics.

Finally, regarding the underlying framework, the model employs a standard Transformer encoder architecture adapted for strain tensor inputs. Detailed hyperparameters, layer configurations, and the schematic diagram are provided in Supplementary Figure S4. Crucially, unlike classical models that rely on engineered descriptors, this architecture enables the model to learn non-linear couplings among strain components directly from data via the self-attention mechanism.

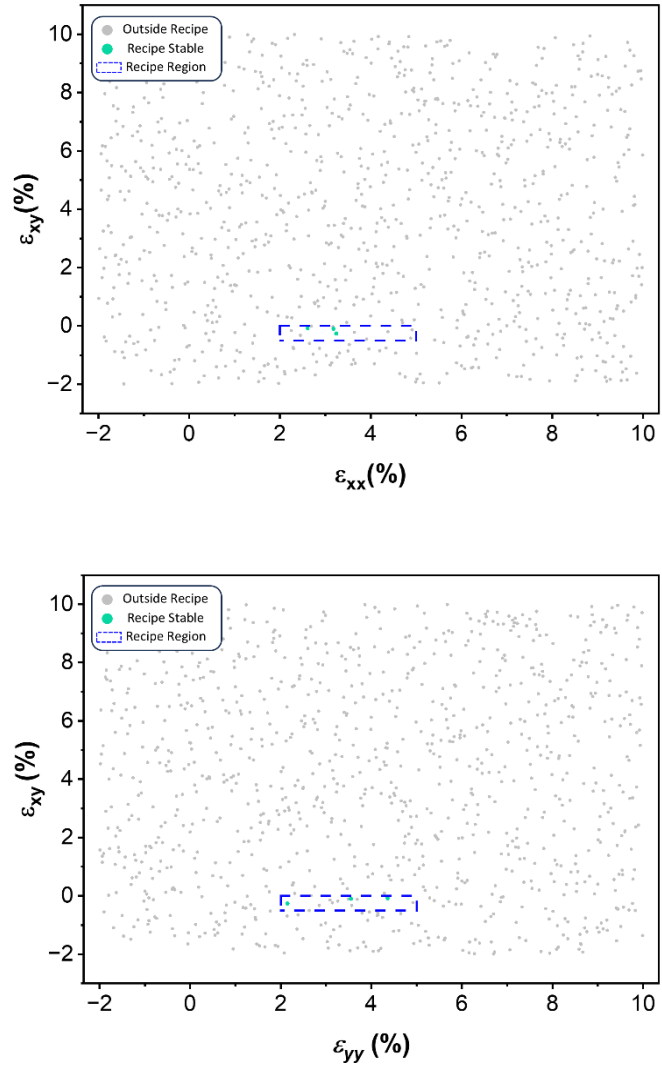

**Figure 5. Validated strain-engineering recipe and stability windows**
Validated strain "recipe" highlighting the 97.7% success and 90.7% stability window $(\varepsilon_{xx}, \varepsilon_{yy} \in [2\%, 5\%]; \varepsilon_{xy} \approx 0\%)$

## Validated Strain Recipes for Practical Applications

To translate the high-throughput screening results into actionable guidance, Figure 5 identifies a validated parameter space where predictions align strictly with computational and physical constraints. Within the moderate biaxial tension window of $\varepsilon_{xx}, \varepsilon_{yy} \in [2\%, 5\%]$ with near-zero shear ($\varepsilon_{xy} \in [-0.5\%, 0\%]$), the framework achieves a 97.7% recipe success rate and a 90.7% stability probability, consistently producing target bandgaps between 4.2 and 4.6 eV.

Crucially, this specific operating window serves as a critical physical validation of our attention-based analysis: while normal strains ($\varepsilon_{xx}, \varepsilon_{yy}$) are effectively used to tune the electronic energy levels, the confinement to near-zero shear confirms that minimizing $\varepsilon_{xy}$ is essential to preserve lattice integrity against symmetry-breaking instabilities. By strictly defining these boundaries, the framework filters out the thermodynamic "dead zones" that typically plague random search strategies. Consequently, this data-driven roadmap significantly streamlines the design process, allowing experimentalists to bypass exhaustive trial-and-error and focus validation efforts exclusively on configurations that are both electronically desirable and physically realizable.

**Conclusion:**

In this study, we demonstrated that a Transformer-based surrogate model can effectively explore the multidimensional strain space of 2D materials, achieving DFT-level accuracy (MAE = 0.0103 eV for bandgap) while eliminating the need for computationally expensive volumetric band-structure inputs. Crucially, beyond numerical prediction, the model's self-attention mechanism served as a physical diagnostic tool: it identified shear strain ($\varepsilon_{xy}$) as the primary driver of symmetry-breaking instabilities, functioning as the "interaction hub" that classical feature metrics failed to capture.

Guided by this insight, we established a validated "safe operation" window ($\varepsilon_{xx}, \varepsilon_{yy} \in [2\%, 5\%]$; $\varepsilon_{xy} \approx 0\%$) that effectively filters out thermodynamic dead zones. This regime guarantees a 97.7% prediction success rate and 90.7% phonon stability, ensuring that the identified configurations are thermodynamically realizable. Ultimately, this work establishes attention-based architectures as powerful dual-purpose tools—capable of both rapid property screening and autonomous physical hypothesis generation—offering a generalized blueprint for accelerating deep elastic strain engineering by bridging the gap between data-driven discovery and physical explainability.

**Methods:**

We focus on hexagonal boron nitride (h-BN) as a prototypical 2D material for strain engineering studies. h-BN exhibits a wide bandgap of ~4.7 eV[23], high thermal stability[24], and strong mechanical properties[25], making it an excellent platform for investigating strain-induced electronic property modulation. Its well-characterized structure and strain-dependent bandgap tunability[23,26] render it particularly suitable for validating our Transformer-based prediction framework, while its relevance to optoelectronic applications[27] provides practical motivation for high-throughput strain engineering.

To construct the dataset, we employed Latin Hypercube Sampling[28] to tile $\varepsilon_{xx}$, $\varepsilon_{yy}$, and $\varepsilon_{xy}$ uniformly over -2% to 10%, while fixing the out-of-plane components to zero due to 2D constraints. Because monolayer materials tolerate tension more readily than compression[29], the design intentionally skews toward tensile strains while retaining a modest compressive range to probe instability. Applying the maximin criterion yields up to 1,000 well-separated samples, providing broad yet efficient coverage of the strain space.

DFT calculations were performed using the Vienna Ab initio Simulation Package[30,31] with the Perdew-Burke-Ernzerhof exchange-correlation functional[32]. All calculations used a 600-eV energy cutoff and a 6×6×1 k-point mesh for electronic structure calculations. Full structural relaxation was performed, allowing both cell shape and volume to vary(Detailed computational parameters, including convergence criteria and vacuum settings, are provided in Supplementary Information Section 1.). Phonon calculations employed a supercell approach, as implemented in the Phonopy code[33,34], where the supercell contains 4×4×1 primitive cells and the corresponding q-point sampling uses a 9×9×1 grid to assess material stability.[35]

Our Transformer model features a 6-dimensional input layer for strain components, which is mapped into a sequence of tokens to maintain a general Voigt tensor representation, followed by a 128-dimensional embedding layer. The architecture includes 8-head multi-head attention, 4 encoder layers, and a 256-dimensional FNN, totaling 209k parameters (Comprehensive hyperparameter configurations for both the Transformer and baseline models are detailed in Supplementary Information Section 2). The self-attention mechanism[18] enables automatic learning of feature interactions through the attention equation:

$$\text{Attention}(Q, K, V) = \text{softmax}\left(\frac{QK^T}{\sqrt{d_K}}\right)V \tag{1}$$

where Q (query), K (key), and V (value) are learned linear transformations of the embedded strain components. The attention weight between strain components $i$ and $j$ is given by:

$$\text{weight}_{ij} = \text{softmax}_j\left(\frac{Q_i K_j^T}{\sqrt{d_K}}\right) \tag{2}$$

where the softmax is applied along the key dimension $j$. In this formulation, $Q_i$ specifies the information required by component $i$, $K_j$ encodes the identity of component $j$, and $V_j$ carries its numerical contribution. The output representation of component $i$ is computed as $\sum_j \text{weight}_{ij} V_j$. Because the mechanism is self-attentive, all strain components function simultaneously as queries, keys, and values. This allows the model to represent potential coupling between strain components, such as interactions between shear strain $\varepsilon_{xy}$ and normal strains $\varepsilon_{xx}$ or $\varepsilon_{yy}$, if such dependencies are present in the data. The attention weights quantify the strength of these interactions, providing interpretable insights into strain-property relationships by revealing, for example, how shear strain $\varepsilon_{xy}$ modulates the influence of normal strains $\varepsilon_{xx}$ and $\varepsilon_{yy}$ on bandgap.

For benchmarking, we compared our Transformer approach against classical ML methods, including XGBoost[36], Random Forest[37], Support Vector Regression (SVR)[38], and FNN. All methods were trained on the same dataset and evaluated using identical metrics to ensure fair comparison of performance and computational efficiency.

Optimization strategies were tailored to each architecture. FNN adopted the Adam optimizer[39] with a learning rate of 0.001, while the Transformer model employed AdamW[40] (an improved variant of Adam that decouples weight decay from gradient updates) with a learning rate of $10^{-4}$. A dropout rate of 0.1 was applied to prevent overfitting. The dataset was split according to model requirements: classical ML methods (XGBoost, Random Forest, SVR) had an 80/20 split, while FNN and Transformer had an 80/10/10 split to enable model selection and early stopping (training history and convergence curves are shown in Supplementary Figure S5). Multi-target prediction included bandgap, direct gap, phonon stability, and strain energy density. Performance was evaluated using Mean Absolute Error (MAE), Root Mean Square Error (RMSE), and coefficient of determination ($R^2$) across all target properties.

**Data availability:**

The datasets generated during and/or analyzed during the current study are available from the corresponding author on reasonable request. Upon acceptance, the complete dataset will be made publicly available on GitHub upon publication.

**Code availability:**

The custom Python code developed for the Transformer-based surrogate model training and attention analysis is available at a GitHub repository to be provided upon publication.

**Acknowledgments**

This work was supported by the National Natural Science Foundation of China (Grant No. 12304098).

**Author contributions**

D.W. conceived the original concept and supervised the project. H.M. designed the methodology, developed the Transformer code, performed the DFT calculations, and wrote the original draft. Y.Z. assisted with data visualization and prepared the schematic workflow diagrams. All authors discussed the results and commented on the manuscript.

**Competing interest**

The authors declare no competing interests.

**Additional information**

Supplementary information is available for this paper. Correspondence and requests for materials should be addressed to D.W